\documentclass[sigconf]{aamas} 

\usepackage{multirow}
\usepackage{hyperref}

\usepackage{balance} 
\usepackage{xcolor}
\usepackage{tikz}
\usepackage{listings}
\usepackage{tabularx, array}
\definecolor{airforceblue}{rgb}{0.36, 0.54, 0.66}
\definecolor{aliceblue}{rgb}{0.94, 0.97, 1.0}
\definecolor{naplesyellow}{rgb}{0.98, 0.85, 0.37}
\definecolor{peach-orange}{rgb}{1.0, 0.8, 0.6}
\definecolor{mulberry}{rgb}{0.77, 0.29, 0.55}
\definecolor{olivine}{rgb}{0.6, 0.73, 0.45}
\usetikzlibrary{positioning,calc,arrows.meta,fit}

\definecolor{applegreen}{rgb}{0.55, 0.71, 0.0}
\usetikzlibrary{shapes.multipart}
\usetikzlibrary{shadings}

\tikzset{
mynode/.style={
  rectangle split,
  rectangle split parts=2,
  rectangle split horizontal,
  draw,
  rectangle split draw splits=false
  }
} 

\definecolor{codegreen}{rgb}{0,0.6,0}
\definecolor{codegray}{rgb}{0.5,0.5,0.5}
\definecolor{codepurple}{rgb}{0.58,0,0.82}
\definecolor{backcolour}{rgb}{0.95,0.95,0.92}

\lstdefinestyle{mystyle}{
    backgroundcolor=\color{backcolour},   
    commentstyle=\color{codegreen},
    keywordstyle=\color{magenta},
    numberstyle=\tiny\color{codegray},
    stringstyle=\color{codepurple},
    basicstyle=\ttfamily\footnotesize,
    breakatwhitespace=false,         
    breaklines=true,                 
    captionpos=b,                    
    keepspaces=true,                 
    numbers=left,                    
    numbersep=5pt,                  
    showspaces=false,                
    showstringspaces=false,
    showtabs=false,                  
    tabsize=2
}

\setcopyright{ifaamas}
\acmConference[AAMAS '26]{Proc.\@ of the 25th International Conference
on Autonomous Agents and Multiagent Systems (AAMAS 2026)}{May 25 -- 29, 2026}
{Paphos, Cyprus}{C.~Amato, L.~Dennis, V.~Mascardi, J.~Thangarajah (eds.)}
\copyrightyear{2026}
\acmYear{2026}
\acmDOI{}
\acmPrice{}
\acmISBN{}

\acmSubmissionID{<<submission id>>}

\title[BMC4TimeSec: Verification Of Timed Security Protocols (Demo)]{BMC4TimeSec: Verification Of Timed Security Protocols (Demo)}

\author{Agnieszka M. Zbrzezny}
\affiliation{
  \institution{Department of Computer Science, Faculty of Design, SWPS University,}
  \city{Warsaw}
  \country{Poland}}
\email{azbrzezny@swps.edu.pl}
 
\begin{abstract}
We present BMC4TimeSec, an end-to-end tool for verifying Timed Security Protocols (TSP) based on SMT-based bounded model checking and multi-agent modelling in the form of Timed Interpreted Systems (TIS) and Timed Interleaved Interpreted Systems (TIIS).

In BMC4TimeSec, TSP executions implement the TIS/TIIS environment (join actions, interleaving, delays, lifetimes), and knowledge automata implement the agents (evolution of participant knowledge, including the intruder). 

The code is publicly available on \href{https://github.com/agazbrzezny/BMC4TimeSec}{GitHub}, as is a \href{https://youtu.be/aNybKz6HwdA}{video} demonstration.

\end{abstract}


\newcommand{\BibTeX}{\rm B\kern-.05em{\sc i\kern-.025em b}\kern-.08em\TeX}

\begin{document}


\pagestyle{fancy}
\fancyhead{}


\maketitle 


\section{Introduction}

Timed Security Protocols (TSPs) utilise network latency, tickets or timestamps, and expiration constraints. The security of these protocols depends on quantitative temporal relationships, including freshness and replay resistance. Conventional verification methods often neglect or oversimplify temporal factors, which hinders the analysis of attacks that exploit time windows, session interleaving, and artefact lifetimes.

BMC4TimeSec addresses these vulnerabilities by integrating the following components:
(i) Multi-agent semantics, based on interpreted systems,
(ii) Interleaving  Semantics (IS) \cite{LomuscioPQ10},
(iii) Satisfiability Modulo Theories-based bounded model checking (SMT-BMC) \cite{Biere09,app142210333} as a practical method for identifying counterexamples to trace length constraints.

The tool offers an end-to-end demonstration, encompassing the entire process from protocol specification to the generation of an interpretable witness.

\paragraph{Relationship to previous work.}
VerSecTis \cite{ZbrzeznyZSSK20,versectis-url} demonstrated agent-based verification of security protocols using SMT.

Subsequently, the paper \cite{app142210333} introduced the first comprehensive approach to Timed Interpreted Systems (TIS) \cite{Wozna-Szczesniak16} with dense time, combined with the SMT-BMC method, and included experimental evaluation on time protocols. BMC4TimeSec extends this works by incorporating TIIS and a modular separation between environment and agent, emphasising interactive analysis and improved usability for users verifying custom protocols, such as Alice-Bob scenarios with JSON-based interpretations.

\section{Agent Technology and Techniques}
BMC4TimeSec adopts a TIIS perspective, where the global state is a composite of the local agent states and the environment state, and evolution occurs within interleaving semantics.

\noindent\textbf{The environment} is represented by \emph{execution automata} generated for protocol sessions/runs.

They control: step order (interleaving multiple sessions), step delays (minimum transfer/execution times), ticket/timestamp clocks and their lifetimes (validity conditions).

\noindent\textbf{Agents} are represented by \emph{knowledge automata} $K(a,t)$ (for participants and the intruder),
which encode whether agent $a$ knows the term/message $t$ and when it acquires this knowledge.

For the intruder, we use \emph{gating}: certain actions are allowed only after knowledge requirements are met, which models the feasibility of constructing messages in the Dolev-Yao model.

\section{Pipeline: from Alice-Bob and JSON to SMT Witness}
The input consists of:
(i) a protocol description in Alice-Bob notation,
(ii) a JSON file with \emph{interpretations} (scenarios), i.e., partial step overrides
(e.g., spoofing, sender/recipient switching, deadline substitutions, intruder message injection),
(iii) a session number parameter $k$.

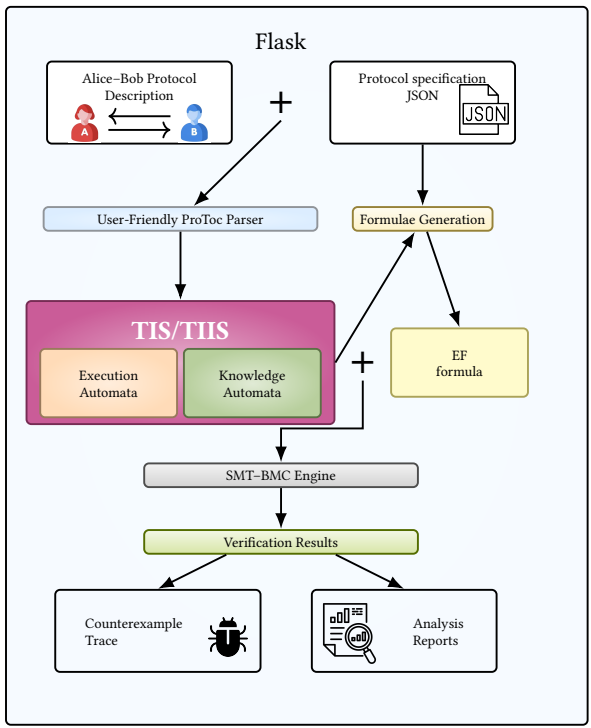
\begin{figure}[t]
\centering
\resizebox{.44\textwidth}{!}{%
\begin{tikzpicture}[
  font=\scriptsize,
  colTIIS/.style={fill=violet!28, draw=violet!55!black},
  colEF/.style={fill=yellow!25, draw=yellow!60!black},
  colFG/.style={fill=yellow!18, draw=yellow!60!black},
  colParser/.style={fill=blue!10, draw=blue!50!black},
  colRes/.style={fill=green!20, draw=green!55!black},
  colSMT/.style={fill=black!12, draw=black!60},
  colK/.style={fill=green!18, draw=green!50!black},
  colE/.style={fill=orange!18, draw=orange!55!black},
  box/.style={
    draw, rounded corners=2pt, thick, align=center,
    minimum height=12mm, minimum width=27mm,
    text depth=-2ex,
    text height=3ex,
    anchor=north,
  },
  wide/.style={box, minimum width=40mm, minimum height=1mm, text depth=-2ex,
    text height=1ex,},
  tiny/.style={box, minimum width=18mm, minimum height=5mm},
  plus/.style={font=\Huge\bfseries},
  arr/.style={-Latex, thick},
]

\node[box, shading=radial,
  inner color=aliceblue!55,
  outer color=aliceblue!60, draw, minimum width=85mm, minimum height=105mm, xshift=23mm, yshift=8mm] (flask) {};

\node[box, inner color=white] (ab) {Alice--Bob Protocol\\Description\\};

\node[draw=none] (Ai) at ([xshift=-8mm, yshift=-3mm]ab) {\includegraphics[scale=0.03]{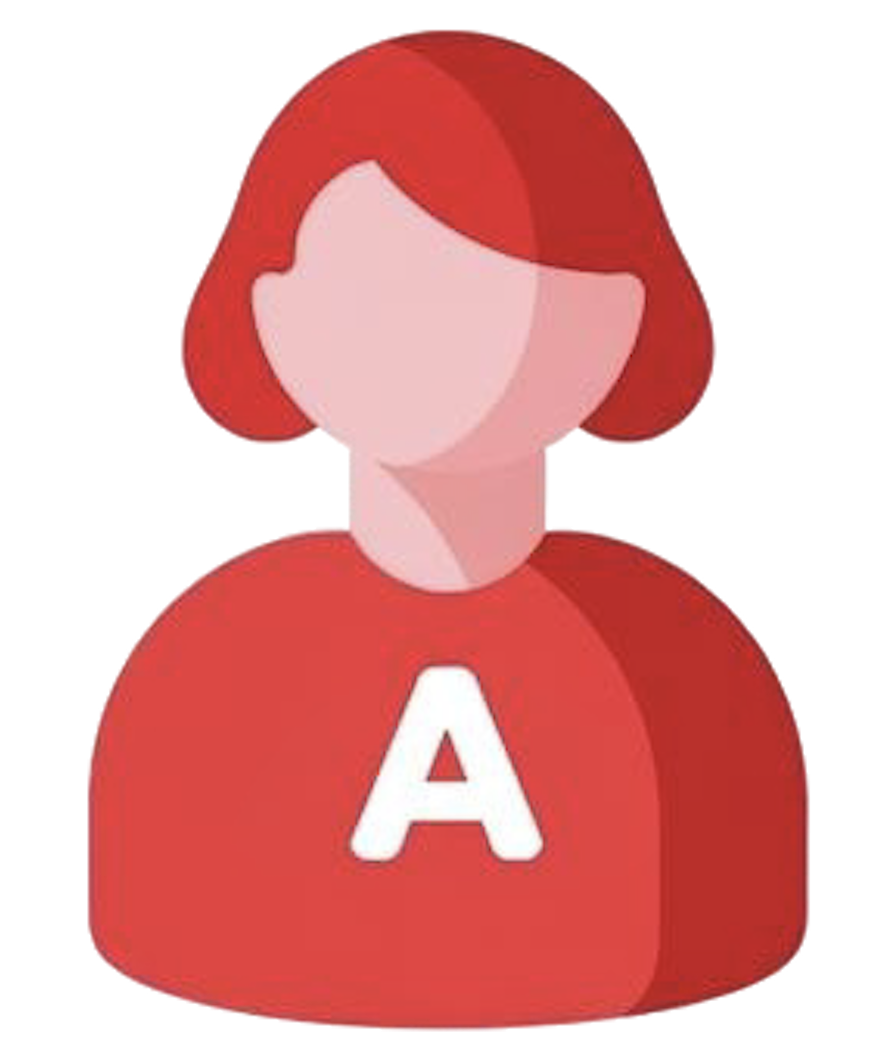}};

\node[draw=none] (Bi) at ([xshift=8mm, yshift=-3mm]ab) {\includegraphics[scale=0.03]{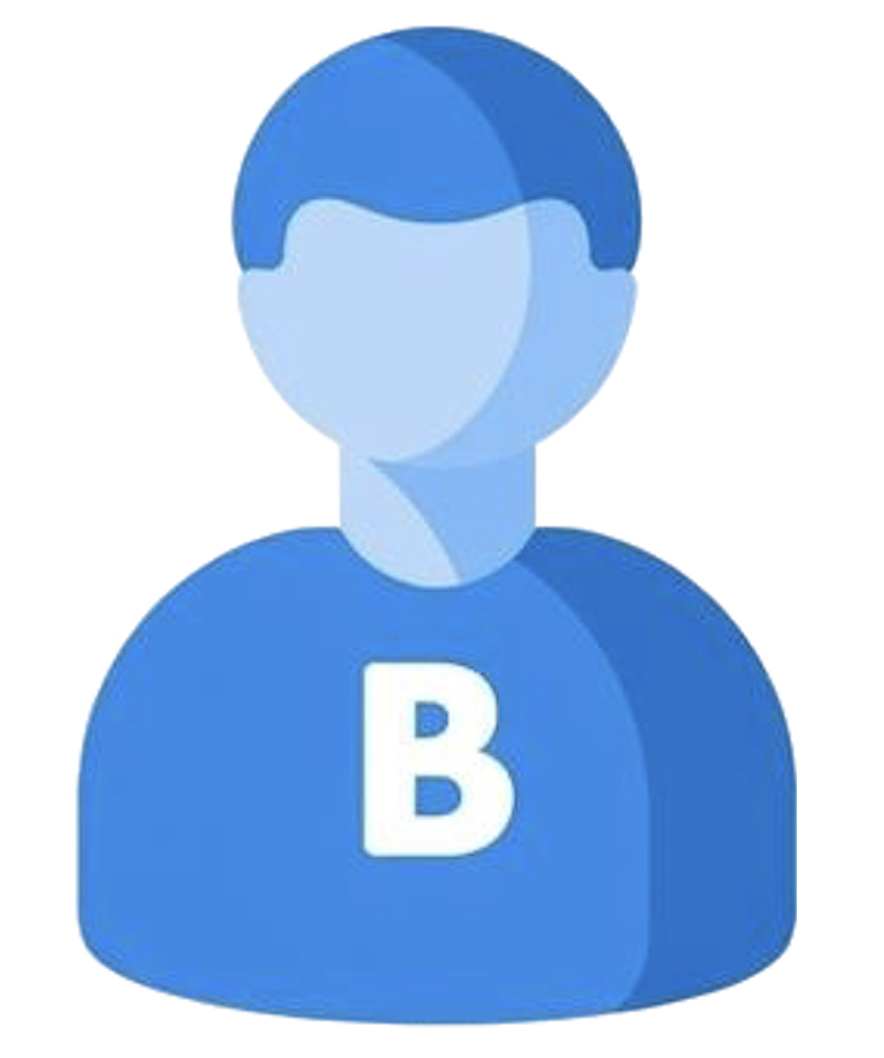}};

\draw[->, thick] ($(Ai.east)+(0,-1mm)$) -- ($(Bi.west)+(0,-1mm)$);
\draw[->, thick] ($(Bi.west)+(0,+1mm)$) -- ($(Ai.east)+(0,+1mm)$);

\node[plus, right=4mm of ab] (p1) {$+$};
\node[above=4mm of p1] (flask_nap) {\large{Flask}};
\node[box, inner color=white, right=4mm of p1,text depth=3.5ex,] (json) {Protocol specification\\ JSON};

\node[draw=none] (js) at ([yshift=-1mm,xshift=9mm]json) {\includegraphics[scale=0.05]{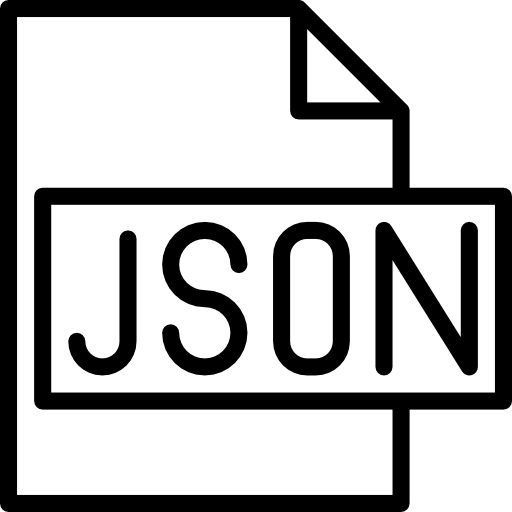}};

\node[wide, colParser, below=9mm of ab, shading=radial,
  inner color=aliceblue!85,
  outer color=aliceblue!230, xshift=6mm, ,draw=aliceblue!60!black] (parser) {User-Friendly ProToc Parser};

\draw[arr] (p1.south) -- (parser);

\node[box, colTIIS, below=10mm of parser, minimum width=45mm, minimum height=18mm, text depth=5ex, text height=0ex, shading=radial,
  inner color=mulberry!45,
  outer color=mulberry!90,draw=mulberry!60!black] (tiis) {\Large\textbf{\textcolor{white}{TIS/TIIS}}};

\node[box, colK, minimum width=20mm, minimum height=10mm, anchor=east, shading=radial,
  inner color=olivine!30,
  outer color=olivine!70,draw=olivine!60!black]
  (know) at ([xshift=-2mm,yshift=-3mm]tiis.east) {Knowledge\\Automata};

\node[box, colE, minimum width=20mm, minimum height=10mm, anchor=west, shading=radial,
  inner color=peach-orange!35,
  outer color=peach-orange!70,draw=peach-orange!60!black]
  (exec) at ([xshift=2mm,yshift=-3mm]tiis.west) {Execution\\Automata};

\draw[arr] (parser) -- (tiis.north);

\node[box, colEF, right=8mm of tiis, minimum width=20mm, minimum height=10mm] (ef) {EF\\formula};

\node[wide, colFG, below=9mm of json,  minimum width=7mm, shading=axis, shading angle=90,
  top color=naplesyellow!10,
  bottom color=naplesyellow!35,draw=naplesyellow!60!black,] (fg) {Formulae Generation};

\draw[arr] (json.south) -- (fg.north);
\draw[arr] (tiis.east) -- (fg);

\draw[arr] (fg) -- (ef.north);

\node[plus] (p2) at ($(tiis.east)!0.5!(ef.west)$) {$+$};

\node[wide, colSMT, below=50mm of p1, shading=axis, shading angle=90,
  top color=gray!10,
  bottom color=gray!35,draw=gray!60!black,] (smt) {SMT--BMC Engine};

\coordinate (join) at ($(p2)+(0,-3mm)$);

\draw[arr] (p2.south) -- ++(0,-7mm) -|(smt.north);

\node[wide, colRes, below=6mm of smt, shading=axis, shading angle=90,
  top color=applegreen!10,
  bottom color=applegreen!35,draw=applegreen!60!black,] (res) {Verification Results};
\draw[arr] (smt) -- (res);

\node[box, below=5mm of res, minimum width=30mm, xshift=-18mm, align=left] (cex) {\hspace{-7mm}Counterexample\\\hspace{-7mm}Trace};

\node[below=8mm of res, xshift=-7.5mm, align=left] (bug) {\includegraphics[scale=0.04]{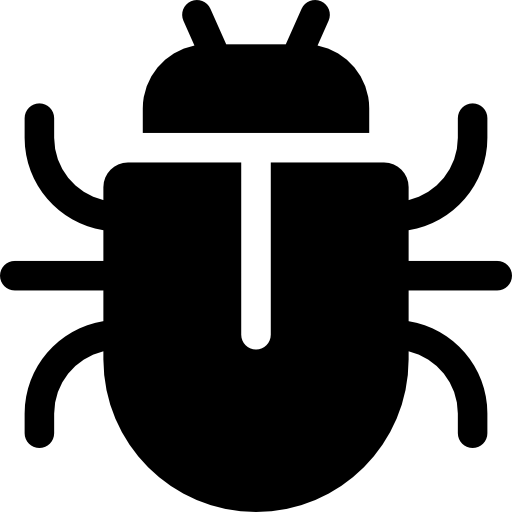}
};

\node[box, draw, below=5mm of res, align=left, xshift=18mm] (rep) {%
  \hspace{10mm}Analysis\\\hspace{10mm}Reports
};

\node[below=6mm of res, xshift=10mm] (icon_res) {\includegraphics[scale=0.065]{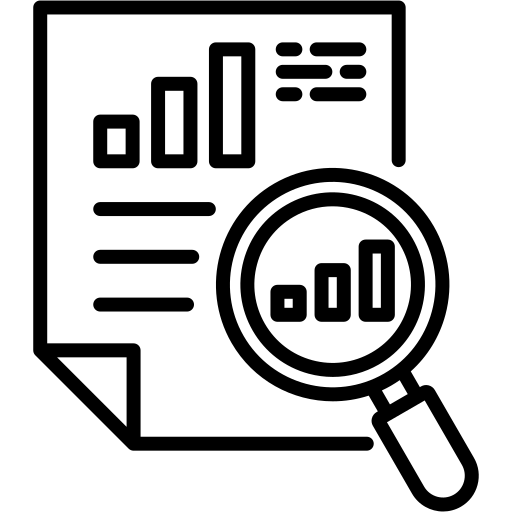}
};

\draw[arr] ($(res.south)+(-8mm,0)$) -- (cex.north);

\draw[arr] ($(res.south) +(8mm,0)$) -- (rep.north);

\end{tikzpicture}
}
\caption{Architecture of BMC4TimeSec.}
\label{fig:bmc4timesec-arch}
\end{figure}

The pipeline is as follows:
\begin{enumerate}
\item \textbf{Generate executions} for multiple sessions and instantiate interpretations (fair and attack).
\item \textbf{Generate a model} with step delays and lifetimes.
\item \textbf{Generating SMT-BMC formulas} for reachability properties, e.g., $EF(\psi)$,
where $\psi$ combines session termination conditions with knowledge conditions (e.g., $K(I,secret)$).
\item \textbf{Running SMT} (e.g., Z3 \cite{MouraB08} in SMT-LIB2) and \textbf{Witness extraction} in the case of SAT.
\item \textbf{Witness analysis}: visualizing TIS/TIIS -- knowledge changes step-by-step. \end{enumerate}
Figure \ref{fig:bmc4timesec-arch} presents the architecture of the tool.

\paragraph{Technology (software stack).}
BMC4TimeSec is a tool implementation (pipeline) that includes: an Alice-Bob parser/specification (implemented in Python), a TIIS generator (implemented in Python), an SMT-BMC formula generators (SMT-LIB2) (implemented in C++), an SMT solver launcher (Z3), and a GUI application (implemented in Python with Flask library).

The demonstration mode emphasises the repeatability of the experiment: the same protocol can be analysed in multiple
interpretations (JSON) and for an increasing number of $k$ sessions.

\subsection{System Innovations}
\paragraph{Comprehensive implementation of Kurkowski-Penczek's idea \cite{KurPen12}}
To the best of our knowledge, BMC4TimeSec is the first such comprehensive and extended implementation of this line:
it integrates (a) time-based execution modelling, (b) agent knowledge modelling, (c) multi-session,
(d) SMT-BMC for TIIS, and (e) counterexample interpretation tools (witness GUI).

\paragraph{Usability: ``anyone can verify their protocol.''}
We include a set of \emph{partial interpretations} (scenarios) for many reference protocols, but the tool is not limited to a library: the user can verify their own protocol by providing an Alice-Bob description and JSON with interpretations. Scenarios can be added without modifying the generator code.

\paragraph{User-friendly specification: ProToc in JSON.}
An important element of usability is the specification language. Based on ProToc~\cite{GrosserKPS14}, we have introduced simplifications and standardisations and converted the specification to JSON format (presented on listing \ref{protoc}), making it easier for users to add interpretations and runtime variants and to support automatic processing by accompanying tools.

\begin{lstlisting}[language=Python, caption=Protoc JSON example,label={protoc}]
    {
      "name": "mitm1_lowe",
      "overrides": [
        {
          "sid": 1,
          "step": 1,
          "kind": "replace",
          "edge": "A->I",
          "L": "<KB,Ta#1|A>"
        },
        {
          "sid": 1,
          "step": 2,
          "kind": "intruder",
          "edge": "I->A",
          "L": "<KA,Ta#1|Tb#1>"
        },
        {
          "sid": 1,
          "step": 3,
          "kind": "replace",
          "edge": "A->I",
          "L": "<KB,Tb#1>"
        },
        {
          "sid": 2,
          "step": 1,
          "kind": "intruder",
          "edge": "I->B",
          "L": "<KB,Ta#1|A>"
        },
        ...
      ]
    }
\end{lstlisting}

\paragraph{More attacks and a broader scope of verification.}
We have expanded the library of attack scenarios compared to previous work: we are not limited to basic spoofing variants or single replays, but also support multi-session and time-dependent scenarios that reveal errors only when sessions are interleaved and lifetime constraints are applied. In practice, this means the ability to automatically analyse, amongst others:
(i) impersonation and man-in-the-middle at the step level, (ii) message or component replay between sessions, (iii) mix-up/mismatch attacks, (iv) non-injective authentication violations (the same freshness element used in multiple sessions), and (v) long-term key compromise scenarios. 
This allows BMC4TimeSec to verify \textbf{more than in previous works}.

\paragraph{Specification Library (TSP)}
With the tool, we provide specifications (Alice--Bob + JSON) for time-based versions of Needham Schroeder Public Key Protocol (NSPK) \cite{NeedhamS78} and its Lowe's modification \cite{Lowe95,SSLK18onSome}; Wide Mouthed Frog Protocol (WMF) \cite{Burrows1989, ZbrzeznySK19} its Lowe's modification of WMF \cite{Lowe97afamily}; the Denning-Sacco Protocol \cite{DenningS81}; the  Kao-Chow Protocol \cite{Kao1995,Szymoniak18}; the  Carlsen's Secret Key Initiator Protocol  \cite{Carlsen94};  the  Needham Schroeder Symmetric Key Protocol (NSSK) \cite{NeedhamS78,ZbrzeznySK19};  the  Yahalom Protocol \cite{Burrows1989}, and its  Lowe's modification \cite{Lowe98}, the  Paulson's modification  (Y$_{P}$) \cite{Paulson01}, and the BAN simplified version  \cite{Burrows1989};  the  Woo Lam Pi Protocol (WLP) \cite{WL94less,ZbrzeznySK19}; the WLP 1, 2, and 3 \cite{WL94less}; the  Andrew Protocol  \cite{S1989IntegratingSI} and its Lowe's modification \cite{Lowe96};  MobInfoSec \cite{SLKFP15ver,SLSK19fast}, and SNEP Protocol  \cite{sensors2021}.
    
For each protocol, we include a fair variant and a set of attack scenarios, including new attack classes that were not covered by VerSecTis.

\begin{acks}
The author gratefully acknowledges the late Prof. Mirek Kurkowski for the inspiration behind this work and for his support and kindness.
\end{acks}



\bibliographystyle{ACM-Reference-Format} 
\bibliography{sample}


\end{document}